\documentclass[10pt]{article}
\usepackage{amssymb}
\newtheorem{proposition}{Proposition}

\def\proofbox{\rule{5pt}{5pt}}

\newcommand{\squar}{\kern1pt\vbox{\hrule height 0.9pt
\hbox{\vrule width 0.9pt\hskip 3pt
\vbox{\vskip 6pt}\hskip 3pt\vrule width 0.6pt}\hrule height
0.6pt}\kern1pt}

\def\ethbar{\overline{\eth}}

\def\sigmab{\overline{\sigma}}

\def\obar{\overline{o}}
\def\iotab{\overline{\iota}}
\def\Ybar{\overline{Y}}
\def\alphab{\overline{\alpha}}

\def\xib{\overline{\xi}}

\def\d{\mbox{d}}

\def\phii{\stackrel{(1)}{\phi}}
\def\phio{\stackrel{(0)}{\phi}}
\def\psii{\stackrel{(1)}{\psi}}
\def\psio{\stackrel{(0)}{\psi}}

\font\tenscr=rsfs10 scaled1100
\font\sevenscr=rsfs7 % scaled \magstep1
\font\fivescr=rsfs5 % scaled \magstep1
\skewchar\tenscr='177 \skewchar\sevenscr='177 \skewchar\fivescr='177
\newfam\scrfam
\textfont\scrfam=\tenscr
\scriptfont\scrfam=\sevenscr
\scriptscriptfont\scrfam=\fivescr

\def\scri{{\fam\scrfam I}}
\def\scrn{{\fam\scrfam N}}
\def\scrs{{\fam\scrfam S}}
\def\scrc{{\fam\scrfam C}}

\def\scrz{{\fam\scrfam Z}}

\def\caln{\mathcal{N}}
\def\cals{\mathcal{S}}

\font\SYM=msbm10
\newcommand{\Real}{{\SYM R}}

\newcommand{\Minkowski}{{\SYM M}}

\newfont{\ghpfontii}{wasy12}

\def\mthorn{\mbox{\ghpfontii j}}

\begin{document}

\title{Polyhomogeneity and zero-rest-mass fields with applications to Newman-Penrose constants.}
\author{Juan Antonio Valiente Kroon.\thanks{E-mail address: {\tt j.a.valiente@qmw.ac.uk}} \\
%EndAName
School of Mathematical Sciences,\\
Queen Mary \& Westfield College,\\
Mile End Road, London E1 4NS,\\
United Kingdom.}

\maketitle

\begin{abstract}

A discussion of polyhomogeneity (asymptotic expansions in terms of $1/r$ and $\ln r$) for zero-rest-mass fields and gravity and its relation with the Newman-Penrose (NP) constants is given. It is shown that for spin-$s$ zero-rest-mass fields propagating on Minkowski spacetime, the logarithmic terms in the asymptotic expansion appear naturally if the field does not obey the ``Peeling theorem''. The terms that give rise to the slower fall-off admit a natural interpretation in terms of advanced field. The connection between such fields and the NP constants is also discussed. The case when the background spacetime is curved and polyhomogeneous (in general) is considered. The free fields have to be polyhomogeneous, but the logarithmic terms due to the connection appear at higher powers of $1/r$.
In the case of gravity, it is shown that it is possible to define a new auxiliary field, regular at null infinity, and containing some relevant information on the asymptotic behaviour of the spacetime. This auxiliary zero-rest-mass field ``evaluated at future infinity ($i^+$)'' yields the logarithmic NP constants.

\end{abstract}

\section{Introduction}

The asymptotic behaviour of the class of spinorial fields known as zero-rest-mass fields is closely interwoven with that of the spacetime background on which the test fields propagate. If the background spacetime describes the gravitational field of an isolated body, then one would expect it to be \emph{asymptotically flat} (i.e. like Minkowski as one recedes from the source). This notion of asymptotic flatness is rather ambiguous, and has given place to the precise definitions of \emph{asymptotic simplicity} and \emph{weak asymptotic simplicity}  of Penrose \cite{penroserindlerii}, \cite{hf98}. These definitions require the existence of a conformal factor $\Omega$, that allows us to endow the physical spacetime with a smooth null 3-dimensional boundary (null infinity, $\scri$), to obtain an extended manifold, the unphysical spacetime. Null infinity consists of two disconnected parts: future null infinity $\scri^+$, and past null infinity $\scri^-$. Penrose's contruction is no doubt elegant and has given rise to a huge amount of developments in the area of the asymptotic behaviour of fields (not only the full non-linear gravity, but also the Maxwell field and linear gravity), however, it is still the subject of debate whether or not there exists a class of physically meaningful spacetimes that are asymptotically simple \cite{hf98}. Specific examples do exist, but they are not completely free of objections \cite{jb95} (it has been shown for example that the C-metric has a complete $\scri$ that is punctured at two points).

When solving the equations for a particular field (a zero-rest-mass field, or even the gravitational field), one has always to decide on the asymptotic behaviour one is going to allow the field to have. Penrose \cite{rp65}, and  Newman \& Penrose \cite{np62} assumed that the fields evaluated at the null boundary of the conformally rescaled spacetime should be \emph{regular} (i.e. bounded, and with at least some given number of derivatives). This choice of asymptotic behaviour turns out to be equivalent to assuming that the components of the relevant fields obey the \emph{Peeling theorem}. As it has been discussed elsewhere \cite{javk99b} and will also be discussed further in the present article, the Peeling theorem is also a consistency condition on the type of asymptotic expansion one should use to calculate the fields. The asymptotic expansions using Penrose's Ansatz can be performed fully using power expansions in $1/r$. However, the analysis of the non-linear stability of Minkowski spacetime has shown that the generic asymptotically simple solutions of the Einstein field equations do not peel in the Penrose sense  \cite{ck}. Moreover, there are a number of results that tend to suggest that a less restrictive setting for the asymptotic study is that of \emph{polyhomogeneous expansion}, i.e. series expansions in terms of powers of $1/r$ and $\ln r$ (see for example the analysis of \cite{andersson2} and \cite{xiv}). One of the key objectives of the present article is point out that the $\ln r$ terms in the asymptotic expansions arise in a natural way when one works with fields that do not peel.

A number of results that hold for peeling fields have their counterpart within the realm of non-peeling polyhomogeneous fields \cite{xiv}, \cite{javk98a}, \cite{javk99a}, however the fact that some of the components of the fields studied are not regular at null infinity makes difficult the direct application of some standard lines of argument. Most of our analysis will be carried out by performing ``formal expansions'' of the diverse components of the fields and the spacetime. The adjective formal is included here because we will not be concerned with the convergence of the series expansions. A crucial observation in the present analysis is the fact that although due to the presence of some particular powers of $1/r$ and $\ln r$ some components of the zero-rest-mass field are not defined at $\scri$, the coefficients attached to these expansions are perfectly well defined functions on $\scri$.

One asymptotic property of the zero-rest-mass field and gravity we will be most interested in is the existence of conserved quantities at null infinity \cite{np65}, \cite{np68}, \cite{javk98a}, \cite{javk99a}. These quantities are known as the Newman-Penrose constants of the field. For a field propagating in Minkowski spacetime (e.g. the Maxwell field or linear gravity) there is an infinite hierarchy of Newman-Penrose constants. If the background spacetime is curved (polyhomogeneous or not), then only the first constants of the hierarchy survive. In general there will be only $2s+1$ conserved quantities for an arbitrary spin-$s$ zero-rest-mass field propagating in a curved spacetime.

The NP constants possess several interesting properties. Among them stands the fact that they can be deduced from the ``formal evaluation of the field'' at future time infinity ($i^+$) by means of an integral representation (Kirchoff-d'Adhemar formula) \cite{np68}, \cite{dcr69}, \cite{penroserindleri}, \cite{penroserindlerii}. Another property is that it is possible to identify the invariant transformations of the Lagrangian of the field that give rise to the conserved quantities (inverse of Noether's theorem) \cite{gg70}. This analysis was carried out for peeling fields, and its principal result is that the invariant transformation has support on the boundary of the spacetime (null infinity). This seems to indicate somehow that the constants arise due to the kind of boundary conditions adopted. Due to problems of regularity on $\scri$ these arguments cannot be applied directly to polyhomogeneous fields. One of the objectives of this article is to provide a solution to this issue. This will result in the gain of deeper insight into the structure of the non-peeling fields and in an unified point of view of the non-logarithmic and logarithmic Newman-Penrose constants.

\bigskip

The article is organized as follows: in section 2, there is a brief discussion of the coordinate system and tetrad that will be used. A working knowledge of the Newman-Penrose (NP) formalism will be assumed. Also, a brief discussion on polyhomogeneous expansions is given, however, for further details we refer the reader to \cite{javk98a} and \cite{javk99a}. In section 3, zero-rest-mass fields propagating on Minkowski spacetime are discussed. The emphasis is set on the properties of these fields when the Peeling theorem does not holds. It is shown that if a field is not peeling, then the asymptotic expansions will contain logarithmic terms. It is also shown that the non-peeling components of the field can be interpreted as produced by advanced radiation. With respect to the NP constants, an old result by Newman-Penrose regarding the vanishing of these quantities for retarded fields is recalled. In section 4, there is a discussion on zero-rest-mass fields propagating on a curved spacetime (polyhomogeneous or not). If the spacetime is polyhomogeneous, then the test fields will also be polyhomogeneous, even if they are regular at $\scri$. This is due to the logarithmic terms that appear in the connection. It is shown that the infinite hierarchy of conserved quantities disappears, and conjectured that only the first set of constants survive. Finally, in section 5, the case of the gravitational field is treated. The gravitational spin-2 field of Penrose is discussed. It is shown  that for a generic polyhomogeneous spacetime, this field is not regular at null infinity. An auxiliary field is constructed out of the logarithmic terms of the components of the Weyl tensor. This field is regular at $\scri$, and hence it allows us to use the full machinery developed for peeling fields. In particular, it allows us to give a new derivation of the logarithmic NP constants. It also sheds some light on why the non-logarithmic quantities are not conserved. In the appendix one finds the expressions for a set of advanced and retarded fields of arbitrary spin. This is a direct generalization of the solutions for linear gravity ($s=2$) given by Torrence and Janis \cite{tj66}.

\section{Notation \& conventions}

\subsection{Coordinates and tetrad}
Stewart \cite{stewart} has constructed a coordinate system and a null tetrad  well-adapted to the study of the asymptotic behaviour of the gravitational field. These coordinates and null tetrads are the ones that will be used in the present article. One begins by assuming that there is a one parameter family of null hypersurfaces ($\caln_u$). One can use the parameter $u$ of the family to define a scalar field. On each of the geodesic generators $\gamma _u$
of the null hypersurfaces it is possible to choose an affine parameter $r$. The null hypersurfaces $\caln_u$ intersect $\scri$ in cuts $\cals_u$. On any one of these cuts, say $\cals_0$, we can choose arbitrary coordinates $x^\alpha $, $\alpha=2,3 $. The coordinates $x^\alpha $ can be propagated on $\scri$ by
demanding $x^\alpha =const$ on each generator of $\scri$. In this way one 
has constructed a coordinate system $(u,r,x^\alpha )$ in the neighbourhood
of $\scri$. 

Once the coordinate system has been set, one can proceed to construct a null tetrad. An obvious choice for the first element of the tetrad is the gradient of the null hypersurfaces 

\begin{equation}
l=\mbox{d}u.
\end{equation}
The vector field $l^a$ is tangent to the geodesic generators 
$\gamma _u$ of $\caln_u$. The freedom of the scaling of the affine
parameter of these generators can be used to set

\begin{equation}
l^a=\frac \partial {\partial r}.
\end{equation}

Let us denote the 2-surfaces $u=const$, $r=const$ by $\cals_{u,r}$. It can be
seen that $l^a$ is future pointing and orthogonal to $\cals_{u,r}$%
. There is only one other null direction with the same properties. The
second vector of our tetrad $n^a$ is chosen to be parallel to
this direction. The two other vectors of the tetrad, $m^a$
and $\overline{m}^a$, are constructed so that they span the
tangent space to $\cals_{u,r}$, $\mbox{T}(\cals_{u,r})$. So from this construction one deduces

\begin{equation}
n^a=\frac \partial {\partial u}+Q\frac \partial {\partial
r}+C^\alpha \frac \partial {\partial x^\alpha },
\end{equation}

\begin{equation}
m^a=\xi ^\alpha \frac \partial {\partial x^\alpha },
\end{equation}
so that the contravariant metric tensor is

\begin{equation} 
g^{ab}=\left(  
\begin{array}{cccc} 
0 & 1 & 0 & 0 \\  
1 & 2Q & C^\theta & C^\varphi \\  
0 & C^\theta & -2 \xi^\theta \overline{\xi}^\theta & -\xi^\theta \overline{\xi}^
\varphi - \overline{\xi}^\theta \xi^\varphi \\
0 & C^\varphi & -\xi^\theta \overline{\xi}^
\varphi - \overline{\xi}^\theta \xi^\varphi & -2 \xi^\varphi 
\overline{\xi}^\varphi
\end{array} 
\right) , 
\end{equation} 
Using a spin-boost it is possible to set $\epsilon =\overline{\epsilon }$
everywhere. From the commutator equations we deduce

\begin{equation}
\kappa =\epsilon =0,
\end{equation}

\begin{equation}
\tau =\overline{\pi }=\overline{\alpha }+\beta \,
\end{equation}
and $\rho $ and $\mu $ are real. At each point of the spacetime one can also introduce a dyad of spinors $(o_A, \iota_A)$ with the normalization $o_A\iota^A=1$($l^a=o^A\obar^{A'}$, $n^a=\iota^A\iotab^{A'}$, $m^a=o^A\iotab^{A'}$).

In the case the spacetime is Minkowski (\Minkowski) one has:

\begin{equation}
 \begin{array}{cc}
 Q=-1/2, & C^\alpha=0, \\
 \rho=-1/r, & \alpha=-\beta=-2^{-3/2}\cot \theta, \\
 \mu=-1/2r & 
 \end{array}
\end{equation}
and
\begin{equation}
\xi^\alpha=2^{-1/2} \left(
 \begin{array}{c}
 1 \\
 -i \csc \theta
 \end{array}
 \right)
\end{equation}
and all the remaining spin coefficients equal to zero.

Quantities evaluated in the conformally rescaled (``unphysical'') spacetime will be denoted by a caret ($\widehat{g}_{ab}=\Omega^2g_{ab}$).

\subsection{Asymptotic expansions and polyhomogeneity}
As mentioned before, the classical work on the asymptotics of linear fields and the gravitational field has been done by means of series expansions in powers of $1/r$. In the present work we will be interested in a broader setting, that of polyhomogeneous expansions (power series in $1/r$ and $\ln r$). Hence it will be assumed that all relevant quantities can be expanded in such a way. For a detailed discussion on the way this can be done within the NP framework the reader is referred to references \cite{javk98a}, \cite{javk99a}. 
If the function $f$ is polyhomogeneous, then we will write

\begin{eqnarray}
f&=&\sum_{i=1}f_ir^{-i} \nonumber, \\
&=&\sum_{i=1}\sum_{j=0}^{N_i}f_{ij}r^{-i}\ln^jr,
\end{eqnarray}
where $f_i$ is a polynomial in $z=\ln r$ with coefficients in $C^k($\Real$\times\cals^2)$ ($f_i=f_{i,N_i}\ln^{N_i}r+\cdots+f_{i,0}$). Differentiation with respect to $z=\ln r$ will be denoted by a prime ($'$). The symbol $\#$ will denote the degree of a polynomial in $\ln r$. In particular  $g'=0$ implies $\#g=0$. The zero polynomial has degree $-\infty$ ($\#0=-\infty$) by definition.

 The expansions that we will use in the present article will be formal, in the sense that we will not be concerned with the convergence of the series expansions. This is mainly because so far there is no existence/uniqueness theorem for polyhomogeneous initial data avaible. A way of looking as this is that all the relevant functions can be expanded up to a certain order in terms of powers of $1/r$ and $\ln r$ plus a residue.

\section{Spin-$s$ zero rest-mass fields on Minkowski spacetime}

\subsection{Some generalities}
The starting point of our discussion will be the class of spinor equations known as the \emph{massless free-field equations for spin} $s$ \cite{rp65}. Let $\phi_{AB\cdots L}$ have $2s$ indices and be symmetric, $\phi_{AB\cdots L}=\phi_{(AB\cdots L)}$. The massless free-field equation for spin $s$ is
\begin{equation}
\nabla^{AA'} \phi_{AB \cdots L} =0.
\end{equation}

 Using the spinors $(o^A,\iota^A)$ the dyad components of the field are:
\begin{equation}
\phi_n=\phi_{A\cdots PQ \cdots K} o^A\cdots o^P\iota^Q\cdots\iota^K,
\end{equation}
where $n$ is the number of $\iota_A$'s, and it ranges from $0$ to $2s$. It can be shown that $\phi_n$ has spin weight $s-n$. The equations for a spin-$s$ zero rest-mass field in the NP formalism are \cite{dcr69}:

\begin{eqnarray}
D\phi_n-\ethbar\phi_{n-1}&=&-(n-1)\lambda\phi_{n-2}+n\pi\phi_{n-1} \nonumber \\
&&-(2(n-s)\epsilon+(n-2s-1)\rho)\phi_n-(2s-n)\kappa\phi_{n+1}, \label{NP2}
\end{eqnarray}

\begin{eqnarray}
\Delta \phi_{n-1} - \eth\phi_n&=&(n-1)\nu\phi_{n-2} -(n\mu +2(n-s-1)\gamma)\phi_{n-1} \nonumber \\
&&-(2s-n+1)\tau\phi_n -(n-2s)\sigma\phi_{n+1}, \label{NP1}
\end{eqnarray}
$n=1,\ldots,2s$. $D$ and $\Delta$ are the covariant directional derivatives in the direction of $l^a$ and $n^a$ respectively, and $\eth$ is the ``eth'' differential operator. The conventions of Penrose \& Rindler \cite{penroserindleri} will be followed. If the spacetime is curved, then $\eth$ can be also expressed in terms of a power series:

\begin{equation}
\eth=r^{-1}\eth_1 +r^{-2}\eth_2+\cdots,
\end{equation}
so that $\eth_n f(r)=0$ \cite{javk99a}. If the background spacetime is flat, then only the first term in the expansion is left ($\eth=r^{-1}\eth_1$).

 The spinorial field $\phi_{AB \cdots L}$ can be recovered from its components via

\begin{equation}
\phi_{AB\cdots K}=(-1)^{2s} \left[ \phi_0\iota_A\cdots \iota_K+\sum_{n=0}^{2s-1}\frac{(2s)!(-1)^{n+1}\phi_{n+1}}{(n+1)!(2s-n-1)!}\iota_{(A\cdots}\iota_Po_{Q\cdots}o_{K)} \right],
\end{equation}
where the indices $A\cdots P$ number $(2s-n-1)$.

When $s=1$ the zero-rest-mass field corresponds to the electromagnetic (Maxwell) field, while the case $s=2$ corresponds to linearized gravity (linear perturbations over a fixed background). Zero-rest-mass fields with a half-integer $s$ correspond to fields that require a quantum interpretation (neutrino fields, etc.), and shall not concern us. Zero-rest-mass fields with an $s>2$ propagating on a curved background have to satisfy a consistency condition (Buchdahl constraint \cite{hab87}). Therefore they cannot be completely arbitrary. \emph{The cases $s=1$, and $s=2$ being the ones with more relevance in our analysis, our discussion and proofs will be limited to them.} In many cases, for the sake of conciseness and clarity only the proofs for $s=2$ will be given. 

\subsection{Peeling and polyhomogeneity}

The classical study of the asymptotic behaviour of these fields has been done by assuming that the components of the field can be expanded asymptotically in powers of $1/r$. Closely related to this kind of expansion is the famous result known as the \emph{Peeling theorem}. If the conformally rescaled zero rest-mass field $\widehat{\phi}_{AB\cdots L}$ is \emph{regular} at $\scri$, then the physical components of the field will satisfy 

\begin{equation}
\phi_n=O(r^{-(2s+1-n)}),
\end{equation}
$n=0\ldots2s$. This results holds for both flat and curved spacetimes.

It is perfectly possible to have zero-rest-mass fields with a decay that is slower than the one prescribed by the Peeling theorem. If this is the case, then polyhomogeneous series will be required to obtain asymptotic expansions. It is important to point out here, that the polyhomogeneous expansions that are used are not the most general kind of expansions one could use. Expansions with fractional powers of $1/r$ cannot be directly ruled out. The results on the nonlinear stability of Minkowski spacetime of Christodolou \& Klainerman give fall off of fractional order (e.g. $\Psi_0=O(r^{-7/2})$). However, if one insists on working with integer powers of $1/r$, then the natural way of extending the results is by means of polyhomogeneous expansions. In this article only integer powers of $1/r$ will be considered.

With this idea in mind, one sees that if the fields do not peel, then the polyhomogeneous expansions arise in a natural way.    

\begin{proposition}
If a spin $s$ zero-rest-mass field propagating on an asymptotically flat spacetime has a fall off that is slower than the one that is prescribed by the Peeling theorem, then asymptotic expansions for the field are polyhomogeneous (i.e. they include powers of $\ln r$).
\end{proposition}

\textbf{Proof.} Consider a massless field that falls slower than is prescribed by the Peeling theorem ---$\phi_0 =O(r^{-(2s+1)})$---. Without loss of generality consider the minimal case (i.e. a field with a fall off one power of $1/r$ slower):
\begin{eqnarray}
\phi_0= \sum_{n=2s}\phi_0^n r^{-n},  \label{prop1}\\
\phi_1= \sum_{n=2s}\phi_1^n r^{-n},  \label{prop2}
\end{eqnarray}
where the coefficients in the expansions are allowed to have a dependence on $\ln r$. The equation (\ref{NP2}) for a zero-rest-mass field propagating in an asymptotically flat spacetime with the tetrad choice of section 2 reads,

\begin{equation}
D\phi_1-\ethbar\phi_0=\pi\phi_0+2s\rho\phi_1.
\end{equation}
Expanding this last equation with the help of (\ref{prop1}) and (\ref{prop2}) yields to the leading order ($\pi$ contains no terms of order $1/r$ in an asymptotically flat spacetime):

\begin{equation}
\phi_1^{2s \prime} = \ethbar \phi_0^{2s}.
\end{equation}
This means that $\phi_1^{2s}$ is a polynomial in $\ln r$ of one degree more than $\phi_0^{2s}$. If $\phi_0^{2s}=0$ identically (a peeling field) then no logarithmic terms appear in the expansions. On the other hand, if this coefficient is different from zero, then the expansions contain logarithmic terms, i.e. they are polyhomogeneous.
 
\proofbox
\bigskip

The general conclusion we extract is that in order to study zero-rest-mass fields that fall slower than is prescribed by the Peeling theorem, polyhomogeneous expansions should be used. A similar result is also true in the case of full non-linear gravity. This is also closely connected to the \emph{Outgoing Radiation Condition} of Bondi and Sachs (see, \cite{vii}, \cite{viii}) and also \cite{javk99b}).

The kind of polyhomogeneous fields used in the previous proof will be known as ``minimal''. This is because these spacetimes violate the Peeling theorem in the simplest possible way, the logarithmic terms in the expansions appearing naturally due to the presence of a non-peeling term in the component $\phi_0$. Recall that this component corresponds to the free data on a null hypersurface $\scrn_0$, all the other initial data is given on $\scri$. Therefore if one wants to put any prescription on the fall off of the field, it has to be done directly on $\phi_0$.

Using arguments on the line of the proof, it is not difficult to show that the components of a ``minimal'' non-peeling spin-$s$ field propagating on Minkowski are given by:

\begin{eqnarray}
\phi_0&=&\phi_0^{2s,0}r^{-2s}+\left(\phi_0^{2s+1,1}\ln r +\phi_0^{2s+1,0}\right)r^{-2s-1}+\cdots, \label{phi0} \\
\phi_1&=&\left(\phi_1^{2s,1}\ln r +\phi_1^{2s,0}\right)r^{-2s}+\left(\phi_1^{2s+1,1}\ln r +\phi_1^{2s+1,0}\right)r^{-2s-1}+\cdots, \label{phi1} \\
\phi_2&=&\phi_2^{2s-1,0}r^{-2s+1}+\left(\phi_2^{2s,1}\ln r +\phi_2^{2s,0}\right)r^{-2s}+\cdots, \label{phi2} \\
 &\vdots& \nonumber \\
\phi_{2s}&=&\phi_{2s}^{1,0}r^{-1}+\cdots+\phi_{2s}^{2s-1,0}r^{-2s+1}+\left(\phi_{2s}^{2s,1}\ln r + \phi_{2s}^{2s,0} \right)+\cdots, \label{phi2s}
\end{eqnarray}
where several relations between the different coefficients hold. In particular, the ``offending'' term $\phi_0^{2s,0}$ is a constant of motion ($\dot{\phi}_0^{2s,0}=0$). The case $s=2$ has been worked out in appendix C, as an illustrative example.

The field given by equations (\ref{phi0})-(\ref{phi2s}) is not regular at $\scri$ because the component $\phi_0$ is not bounded there. However let us  consider the auxiliary field 

\begin{eqnarray}
\phii_0&=&\phi_0^{2s+1,1}r^{-2s-1}+\phi_0^{2s+2,1}r^{-2s-2}+\cdots, \\
\phii_1&=&\phi_1^{2s,1}r^{-2s}+\phi_1^{2s+1,1}r^{-2s-1}+\cdots, \\
&\vdots& \nonumber \\
\phii_{2s}&=&\phi_{2s}^{2s,1}r^{-2s}+\phi_{2s}^{2s,1}r^{-2s}+\cdots,
\end{eqnarray}
constructed out of the logarithmic coefficients in equations (\ref{phi0})-(\ref{phi2s}) (i.e. coefficients of the form $\phi_n^{x,1}$). It is not difficult to show that it satisfies formally the zero-rest-mass field equation $\nabla^{AA'}\phii_{AB\cdots L}=0$, and is regular at $\scri$. That this affirmation is true for the case $s=2$ can be seen by comparing the results of the appendices B and C. As will be seen later, this field carries some information about the asymptotic behaviour of the polyhomogeneous field, but not about its radiative properties, as the component $\phii_{2s}$ (generally associated to the radiation field) goes as $1/r^{2s}$, and not as the usual $1/r$ characteristic of radiation. In other words, $\phii_{AB\cdots L}$ describes the ``higher asymptotics'' of the polyhomogeneous zero-rest-mass field $\phi_{AB\cdots L}$.

Similarly, one can construct another spinorial field from the non-logarithmic coefficients in (\ref{phi0})-(\ref{phi2s}), and demand that the series expansion contains no $\ln r$ in it:

\begin{eqnarray}
\phio_0&=& \phi_0^{2s,0}r^{-2s}+\phi_0^{2s+1,0}r^{-2s-1}\cdots, \\
\phio_1&=& \phi_1^{2s,0}r^{-2s}+\phi_1^{2s+1,0}r^{-2s-1}\cdots, \\
 &\vdots& \nonumber \\
\phio_{2s}&=&\phi_{2s}^{1,0}r^{-1}+\phi_{2s}^{2,0}r^{-2}\cdots.
\end{eqnarray}
Due to the requirement that the components of the field $\phio_{AB\cdots L}$ contain no logarithmic terms, it will not satisfy the spin-$s$ zero-rest-mass field equation, but a non-homogeneous version of it:

\begin{equation}
\nabla^{AA'}\phio_{AB\cdots L}=f^{A'}_{B \cdots L}.
\end{equation}
The spinor $f^{A'}_{B \cdots L}$ is completely symmetric on the indices $A,\ldots,L$, and can be expressed in terms of the coefficients of $\phii_0$ and its derivatives. 

\bigskip

It is well known that a general zero-rest-mass field propagating on Minkowski spacetime is a combination of advanced and retarded fields. It is a very difficult problem to decide whether a particular field is retarded or advanced. However, there is a group of solutions to the zero-rest-mass field equations for which this issue is settled. Torrence \& Janis \cite{tj66} found a class of solutions for the spin-2 field on Minkowski that are constants on past null cones $v=u+2r$. This fact allows us to interpret these solutions as representing advanced radiation. Then, by performing a time reversal of the solution, they were able to deduce the  corresponding retarded field solutions. The results of Torrence \& Janis for the spin-2 field can be easily extended for the case of a field of arbitrary spin $s$. The general expressions are found in the appendix A. Using these expressions, it is possible to give an interpretation of the polyhomogeneous zero-rest-mass fields in \Minkowski, in terms of a suitable combination of advanced and mixed radiation fields; the ``non-peeling'' and logarithmic terms in the asymptotic expansions of the field produced by a  particular kind of the Torrence-Janis advanced solutions.

\begin{proposition}
A linear field in \Minkowski \,(Maxwell field, linear gravity) that is polyhomogeneous can be written as the superposition of an advanced and mixed field, where the advanced field (incoming radiation) accounts for the coefficients violating the Peeling theorem.
\end{proposition}

\textbf{Proof.} In order to ease calculations, the calculations will be restricted to the case $s=2$ (linear gravity). Extensions to an arbitrary spin can be easily done.  Also, without loss of generality let the polyhomogeneous field be minimal, i.e.

\begin{equation}
\phi_0=\phi_0^4r^{-4}+\cdots. \label{minimalphi0}
\end{equation}
The coefficient $\phi_0^4$ is a constant of motion (see the previous discussion, and appendix C), hence one can express it in terms of spin 2 spherical harmonics only,

\begin{equation}
\phi_0^4(\theta,\varphi)=\sum \left(\phi_0^4 \right)_{lm}(_2Y_{lm}),
\end{equation}
where the coefficients $\left(\phi_0^4 \right)_{lm}$ are numerical constants. On the other hand, for a purely advanced (incoming) spin-2 zero rest-mass field one has
\begin{equation}
\phi_0=K_2\sum_{l,m}(_2Y_{l,m})r^{l-2}\partial_r^{l+2}\left(\frac{B^{lm}(v)}{r^{l-1}} \right). \label{outgoingphi0}
\end{equation}
An intelligent guess shows that the form of $B^{lm}$ necessary to account for the $1/r^4$ term in $\phi_0$ is
\begin{eqnarray}
B^{lm}&=&B_0^{lm}(u+2r)^{l-1}\ln(u+2r), \\
&=&(2r)^{-l}\{\ln2 + \ln r +(\mbox{terms of order }1/r) \}
\end{eqnarray}
where the coefficient $B_0^{lm}$ is a constant. After some calculations, it is possible to find the specific relation between the coefficients $\left(\phi_0^4 \right)$ and $B_0^{lm}$ 
\begin{equation}
\left(\Psi_0^4 \right)_{lm}=(-1)^{l+1}2^{l-1}(l+1)!K_{2}B_0^{lm}. \label{relation}
\end{equation}
Therefore, an advanced field satisfying (\ref{relation}) can account for the non-peeling coefficient in (\ref{minimalphi0}).
Because of the linearity of the field propagating on \Minkowski, one can account for the higher order terms in $\Psi_0$ by an suitable superposition of more rapidly decaying retarded and advanced (mixed) fields. 
\proofbox
\bigskip

\subsection{The Newman-Penrose constants}
A well-known property of ``peeling'' zero rest-mass fields propagating on Minkowski spacetime is the existence of an infinite  hierarchy of complex conserved quantities associated to the asymptotic expansion of the field (\emph{the Newman-Penrose constants of the spin-s zero rest-mass field}).  The conserved quantities can be readily deduced by taking  equations (\ref{NP1}) and (\ref{NP2}) for $n=1$ in \Minkowski:
\begin{eqnarray}
\partial_r\phi_1-\ethbar\phi_0=-\frac{2s}{r}\phi_1, \\
\dot{\phi}_0-\frac{1}{2}\partial_r\phi_0-\eth\phi_1=\frac{1}{2r}\phi_0.
\end{eqnarray}
If one assumes that the field zero-rest-mass field ``peels'',
\begin{eqnarray}
\phi_0=\sum_{i=2s+1}\phi_0^ir^{-i}, \\
\phi_1=\sum_{i=2s}\phi_1^ir^{-i},
\end{eqnarray}
then after some calculations one obtains

\begin{equation}
(k-2s)\dot{\phi}_0^{k+1}=-\{ \ethbar\eth +k(k-2s-1)\}\phi_0^k, \label{linearnp}
\end{equation}
$k=2s+1,\dots$. From these equations it is not difficult to deduce the following hierarchy of conserved constants for a (peeling) field of spin-$s$ \cite{np68}:
\begin{equation}
\mathcal{Q}_{k,m}^{(s)}= \oint_{\cals^2}\phi_0^{k}(_s\Ybar_{k-s-2,m}) \d \cals, \label{nphierarchy}
\end{equation}
with $k=2s+2,\ldots$ and $-k+s+2 \leq m \leq k-s-2$. In the case the field is non-peeling (i.e. polyhomogeneous) the quantities given by (\ref{nphierarchy}) will not in general be constants, however it is possible to construct an adequate generalization, \emph{the logarithmic Newman-Penrose constants}. This can be done by a direct calculation on the polyhomogeneous expansions, or more elegantly by looking at the Newman-Penrose constants of the auxiliary field $\phii_{ABCD}$. The logarithmic NP constants of $\phi_{ABCD}$ and the NP constants of $\phii_{ABCD}$ are equal by construction. By the same token, one can see why quantities with the same structure of (\ref{nphierarchy}) in $\phio_{ABCD}$ are not conserved: the auxiliary field $\phio_{ABCD}$ does not satisfies the zero-rest-mass field equation!

As was pointed out in early works by Newman \& Penrose \cite{np68} the NP constants are trivial for linear fields propagating in \Minkowski \, in the sense that they vanish for purely retarded fields. Hence, they just provide the asymptotic profile of incoming radiation.

\begin{proposition}
The Newman-Penrose constants of a purely outgoing linear field (e.g. Maxwell field or linear gravity) vanish.
\end{proposition}
 
\textbf{Proof.} We will limit ourselves to the case of a peeling spin-2 field. The component $n=0$ for a purely retarded field is given by (from the appendix):

\begin{equation}
\phi_0=K_{+2}\sum_{l,m}(_2Y_{l,m})r^{l-2}(\partial_r-2\partial_u)^{l-2}\left(\frac{b^{lm}(u)}{r^{l+3}} \right). \label{retardedphi0}
\end{equation}
KirchoffDue to the orthogonality properties of the spherical harmonics, the $k$-order NP constants ($k=2s+2,\ldots$) will be different from zero if the coefficient $\phi_0^k$ (that goes with $1/r^k$) contains $l=k-4$ harmonics. However, a close examination of equation (\ref{retardedphi0}) shows that the required spherical harmonics appear at order $1/r^5$ and $1/r^{k-1}$, not at order $1/r^k$ as required. Therefore the NP constants are zero (the whole hierarchy!).
\proofbox
\bigskip

When the spacetime is curved, the only remaining constants are the first ones in the hierarchy ($k=2s+2$):

\begin{equation}
\mathcal{Q}^{(s)}_m=\mathcal{Q}_{s,m}^{(s)}= \oint_{\cals^2}\phi_0^{2s+2}(_s\Ybar_{s,m}) \d \cals.
\end{equation}
The fact that these constants are the important ones to look at is also strengthen by the fact that these are the ones that can be deduced using the so-called \emph{generalized -d'Adhemar formula}. This formula allows us to know the value of a zero rest-mass field $\phi_{AB\cdots L}$ at a point $P$ in terms of its value at the intersection $\scrs$ of its (say) past light cone $\scrc$ with a null hypersurface $\scrn$ ($\scrs=\scrc\cap\scrn$):

\begin{equation}
\phi_{AB\cdots L}=\frac{(-1)^{2s}}{2\pi}\oint_{\scrs}\frac{1}{p}\iota_A\cdots \iota_L \mthorn_{\scrc}\phi_0 \d \scrs,
\end{equation}
where $\phi=\phi_{AB\cdots L}o^Ao^B \cdots o^L$ is the null datum at $\scrn$, and $\mthorn_{\scrc}=\partial_r-2s(\rho+\epsilon)-\rho$ \cite{penroserindleri}. Let $Q$ be a typical point on the cross section $\scrs$ of the light cone $\scrc$. The quantity $p$ is the measure of the extent of the portion of null geodesic joining the point $Q$ and $P$, scaled by the tetrad vector $n^a$; that is, $p^2 \d \sigma=\d \scrs$, where $\d \Sigma$ is the area element traced out on the tangent space at $P$ by the vector $n^a$.

This integration can also be performed when the light cone $\scrc$ becomes $\scri^+$ so that one could think of measuring the field at $i^+$. The ``value'' of the field  has to be independent of the cut of $\scri^+$ one chooses to integrate, so that in this case one has $2s+1$ linearly independent complex quantities referred to as the \emph{Newman-Penrose} (NP) constants of the spin-s zero rest mass field:

\begin{equation}
\mathcal{Q}^{(s)}_m=\oint_{\cals^2}(_s\Ybar_{s,m})\phi_0^{2s+1} \d \cals.
\end{equation}

\section{Zero-rest-mass field in an asymptotically flat spacetime}

If the zero-rest-mass field is propagating on a curved spacetime, then one cannot talk any longer about purely retarded/advanced fields. This is mainly because one would expect some backscattering of the field with the background, producing incoming radiation if the field is outgoing, and vice versa. Hence a purely retarded field that is let evolve on its own, will eventually develop an advanced component. 

\subsection{If the background peels}
The evolution equations for the diverse components of the field $\phi_{ABCD}$ will be modified  due to the curvature of the spacetime. This will mean that the relations \ref{linearnp} that gives rise to the infinite hierarchy of NP constants are in general no longer valid. Only for the first term of the hierarchy, the curvature due terms seem to ``conspire'' to get conserved quantities. Whether this happens for some (unknown) higher order is still an open question, however, it may be quite unlikely because of the ever growing number of terms.

\subsection{If the background does not peel}
If the background does not peel (polyhomogeneous), then things get much more complicated. It is clear that the presence of $\ln r$ terms in the connection will give rise to $\ln r$ terms in the expansions of the zero-rest-mass field. Moreover, for some polyhomogeneous backgrounds one finds that a field that initially peels (on $\scrn_0$) will generate non-peeling terms. In order to see this, consider the evolution equation for the $\phi_0$ component of a spin 2 zero-rest-mass field:

\begin{equation}
\Delta \phi_0- \eth\phi_1=(4\gamma-\mu)\phi_0 -4\tau\phi_1+3\sigma\phi_2. \label{Bb}
\end{equation}
If the background spacetime is such that $\Psi_0=O(1/r^3)$, then the leading term of the shear (order $1/r^2$) will contain logarithmic terms \cite{javk98a}, \cite{javk99a}. Equation (\ref{Bb}) at order $1/r^5$ yields:

\begin{equation}
\dot{\phi}_0^5-\eth_1\phi_1^4=3\sigma_2\phi_2^3.
\end{equation}
Therefore, if $\sigma_2$ contains $\ln r$ terms then $\dot{\phi}_0^{5,1} \neq0$, and hence the field will develop logarithmic terms in $\phi_0^5$. The resulting zero-rest-mass field will not be regular at $\scri$.

If the background spacetime is minimally polyhomogeneous, the $\ln r$ terms in the field  will appear at order $1/r^6$ ($\sigma_2$ is logarithm-free), and it will peel for later times if it did initially. The evolution equation for the $1/r^6$ coefficient in $\phi_0$ ($\phi_0^6$) is:

\begin{equation}
\dot{\phi}_0^6=-\ethbar\eth\phi_0^5-4\ethbar(\sigma_2\phi_1^2)+3\sigma_3\phi_2^3.
\end{equation}
The logarithmic terms in $\phi_0^6$ appear due to the presence of the term $3\sigma_3\phi_2^3$ in the right-hand side ($\sigma_3=\sigma_{2,1}\ln r + \sigma_{2,0}$ for a minimal non-peeling spacetime \cite{javk99a}). The term $\phi_2^3$ (the leading one in $\phi_2$) is usually interpreted as the monopolar part of the spin 2 field. When it is not present the $\ln r$ terms in the expansions due to the connection will appear at a higher order in $1/r$: $O(r^{-7})$. 

The construction of auxiliary fields for non-peeling zero-rest-mass fields  described in section 3 can be extended to the case of curved spacetimes. For the sake of clarity and conciseness, it will be assumed that both the spin 2 zero-rest-mass field and the spacetime background are minimal. Extensions to for general kind of fields, background and spin can be done in a fairly straightforward way.

 From the previous discussion one knows that the auxiliary fields (being fields propagating in a polyhomogeneous spacetime) will have $\ln r$ terms in their expansions. However, the auxiliary fields have no monopolar part (the leading terms in $\psi_2$), and therefore these logarithmic terms will appear at order $O(r^{-7})$. Hence, the auxiliary field will still be regular at $\scri$.

As mentioned before, the characteristic initial value problem for a zero-rest-mass field is set by giving

\begin{equation}
\phi_0=\phi_0^4r^{-4}+\left( \phi_0^{5,1}\ln r +\phi_0^{5,0} \right)r^{-5}+\left( \phi_0^{6,1}\ln r +\phi_0^{6,0} \right)r^{-6}+\cdots,
\end{equation}
on an initial null hypersurface $\scrn_0$, $\widehat{\phi}_1$ and $\widehat{\phi}_2$ on $\scrz=\scrn_0\cap\scri$, and $\widehat{\phi_4}$ on $\scri^+$ \cite{rp65}. The minimal polyhomogeneous (and non-peeling) field obtained from this initial value problem has the standard form,

\begin{eqnarray} 
&&\phi_0=\phi_0^4r^{-4}+\left( \phi_0^{5,1}\ln r +\phi_0^{5,0} \right)r^{-5}+\left( \phi_0^{6,1}\ln r +\phi_0^{6,0} \right)r^{-6}+O(r^{-7}\ln r),\\
&&\phi_1=\left( \phi_1^{4,1}\ln r +\phi_1^{4,0} \right)r^{-4} +O(r^{-5}\ln r), \\
&&\phi_2=\phi_2^{3,0}r^{-3}+O(r^{-4}\ln r), \\
&&\phi_3=\phi_3^{2,0}r^{-2}+\phi_3^{3,0}r^{-3}+O(r^{-4}\ln r), \\
&&\phi_4=\phi_4^{1,0}r^{-1}+\phi_4^{2,0}r^{-2}+\phi_4^{3,0}r^{-3}+O(r^{-4}\ln r).\end{eqnarray} 

From this field one can construct out of the logarithmic coefficients ($\phi_n^{x,1}$ the auxiliary field

\begin{eqnarray}
&&\phii_0=\phi_0^{5,1}r^{-5}+\phi_0^{6,1}r^{-6}+O(r^{-7}\ln r), \\
&&\phii_1=\phi_1^{4,1}r^{-4}+\phi_1^{5,1}r^{-5}+\phi_1^{6,1}r^{-6}+O(r^{-7} \ln r), \\
&&\phii_2=\phi_2^{4,1}r^{-4}+\phi_2^{5,1}r^{-5}+O(r^{-6} \ln r), \\
&&\phii_3=\phi_3^{4,1}r^{-4}+\phi_3^{5,1}r^{-5}+O(r^{-6} \ln r), \\
&&\phii_4=\phi_4^{4,1}r^{-4}+\phi_4^{5,1}r^{-5}+O(r^{-6} \ln r), \\
\end{eqnarray}
that can be shown to satisfy the spin 2 zero-rest-mass field equation,

\begin{equation}
\nabla^{AA'}\phii_{ABCD}=0.
\end{equation}
If the polyhomogeneous test field is propagating on a background that is also polyhomogeneous, then the higher order components (those contained in the order symbol) will not coincide with those of the logarithmic terms in the original field. This as discussed previously is due to the presence of logarithms in the connection. Therefore, in order to define the auxiliary field correctly, one has to do it from an initial value problem. This auxiliary characteristic initial value problem is set by giving

\begin{equation}
\phii_0=\phi_0^{5,1}r^{-5}+\phi_0^{6,1}r^{-6}+\cdots, \\
\end{equation}
on $\scrn_0$ together with $\widehat{\phii}_1=\phi_1^{4,1}$, $\widehat{\phii}_2=0$ on $\scrz$, and $\widehat{\phii}_3=\widehat{\phii}_4=0$ on $\scri^+$. 

As discussed previously, the auxiliary field contains no monopolar term, and hence the logarithmic terms due to the connection appear at order $1/r^7$. Details of the calculations involved are shown in the appendix D. Hence the leading terms of the auxiliary field (up to order $1/r^6$) will be the same as those in the logarithmic terms of the original field. In particular, the logarithmic NP constants of the field will coincide with the NP constants of the auxiliary field.

\section{The gravitational field}

\subsection{A hierarchy of auxiliary fields}
It is tempting to extend some of the ideas discussed in the previous sections to the case of the full nonlinear gravitational field. All the tools needed are already in our hands, only some precisions will be needed in terms of defining a zero-rest-mass field that describes the asymptotic behaviour of the gravitational field.

A well known property of the Weyl spinor $\Psi_{ABCD}$ is that it is conformally invariant \cite{penroserindlerii}, \cite{stewart}, i.e.

\begin{equation}
\widehat{\Psi}_{ABCD}=\Psi_{ABCD},
\end{equation}
while the Einstein field equations are not conformally invariant (a vacuum solution in the physical spacetime may no be so one the unphysical manifold).
In contrast, a massless spin-2 field $\phi_{ABCD}$ has to transform as
\begin{equation}
\widehat{\phi}_{ABCD}=\Omega^{-1}\phi_{ABCD}
\end{equation}
in order to keep the zero rest-mass field equations conformally invariant. Since the vacuum Bianchi identities are simply the massless spin-2 field equations, it is legitimate to define a spin-2 zero-rest-mass field (\emph{the gravitational spin-2 field})  by \cite{rp65}, \cite{penroserindlerii}

\begin{equation}
\psi_{ABCD}=\Psi_{ABCD},
\end{equation}
so that $\widehat{\psi}_{ABCD}=\Omega^{-1}\psi_{ABCD}=\Omega^{-1}\Psi_{ABCD}$. The different components of $\widehat{\psi}_{ABCD}$ are of relevance in the study of gravitational radiation. When the spacetime ``peels'' several results concerning massless fields can be directly applied to the spinor $\psi_{ABCD}$. However, as discussed before (see also \cite{javk99b}) if the spacetime is polyhomogeneous then the Peeling theorem is no longer valid, and therefore some of the components of  $\widehat{\psi}_{ABCD}$ diverge at $\scri$ ( $\widehat{\psi}_0$, and $\widehat{\psi}_1$ if $\Psi_0=O(r^{-4}\ln^{N_4}r)$, plus $\widehat{\psi}_2$ if the spacetime has $\Psi_0=O(r^{-3}\ln^{N_3}r)$). This clearly precludes the direct application of some results which require the finiteness of the fields at null infinity.
`
One is therefore left again with the necessity of constructing a zero rest-mass field with a regular behaviour at $\scri$, that gives information about the higher asymptotics of the Weyl tensor. 

As before, for clarity and  conciseness the analysis will be limited to the case of minimal polyhomogeneous spacetimes. The extension to more general cases can be done in a fairly straightforward way. The components of the Weyl tensor for a minimal polyhomogeneous spacetime can be expanded as:

\begin{eqnarray}
\Psi_0=\psi_0&=&\Psi_0^{4,0}r^{-4}+\left( \Psi_0^{5,1}\ln r+\Psi_0^{5,0} \right)r^{-5}+\left( \Psi_0^{6,1}\ln r+\Psi_0^{6,0} \right)r^{-6}+O(r^{-7}\ln r), \label{minimal0} \\
\Psi_1=\psi_1&=&\left( \Psi_1^{4,1}\ln r+\Psi_1^{4,0} \right)r^{-4}+\left( \Psi_1^{5,1}\ln r+\Psi_1^{5,0} \right)r^{-5}+O(r^{-6}\ln r), \label{minimal1} \\
\Psi_2=\psi_2&=&\Psi_2^{3,0}r^{-3}+\left( \Psi_2^{4,1}\ln r+\Psi_2^{4,0} \right)r^{-4}+O(r^{-5}\ln r), \label{minimal2}\\
\Psi_3=\psi_3&=&\Psi_3^{2,0}r^{-2}+\Psi_3^{3,0}r^{-3}+\left( \Psi_3^{4,1}\ln r+\Psi_3^{4,0} \right)r^{-4}+O(r^{-5}\ln r), \label{minimal3}\\
\Psi_4=\psi_4&=&\Psi_4^{1,0}r^{-1}+\Psi_4^{2,0}r^{-2}+\Psi_4^{3,0}r^{-3}+\left( \Psi_3^{4,1}\ln r+\Psi_3^{4,0} \right)r^{-4}+O(r^{-5}\ln r)\label{minimal4},
\end{eqnarray}
where
\begin{equation}
\begin{array}{ll}
\dot{\Psi}_0^{4,0}=0, &  \\
\Psi_1^{4,1}=\ethbar\Psi_0^{4,0}, & \Psi_1^{5,1}=-\ethbar \Psi_0^{5,0}, \\
\Psi_2^{4,1}=-\ethbar\Psi_1^{4,1}, & \\
\Psi_3^{2,0}=-\eth \dot{\sigmab}_{2,0}, & \Psi_3^{3,0}=-\ethbar \Psi_2^{3,0},\\
\Psi_4^{1,0}=-\ddot{\sigmab}_{2,0}, & \Psi_4^{2,0}= \ethbar \eth \dot{\sigmab}_{2,0}.
\end{array}
\end{equation}

These series expansions are obtained by solving formally the \emph{asymptotic characteristic initial value problem} \cite{kannar}, which is set by supplying $\Psi_0$ on an initial null hypersurface $\scrn_0$ (i.e. the set of coefficients $\{\Psi_0^{4,0},\, \Psi_0^{5,1}, \, \Psi_0^{5,0}, \ldots \}$ on $\scrn_0$), $\widehat{\Psi}_1$ and $\mbox{Re}\, \widehat{\Psi}_2$ (the coefficients $\Psi_1^{4,0}$ and $\mbox{Re}\, \Psi_2^{3,0}$) on the intersection of the initial null hypersurface and future null infinity $\scrz=\scrn_0 \cap \scri^+$, together with $\widehat{\sigma}$ ($\sigma_{2,0}$) on $\scri^+$. As it has been discussed elsewhere \cite{xiv}, \cite{javk98a}, \cite{javk99a}, the logarithmic terms in $\Psi_0$ are necessary, even if they are not present at $\scrn_0$: the evolution equations, together with the coefficient $\Psi_0^{4,0}$ ---that violates the Peeling theorem--- will give rise to them.

 Hence, define the totally symmetric spinorial field $\psii_{ABCD}$ so that its leading terms are given by,
\begin{eqnarray}
\psii_0&=&\Psi_0^{5,1}r^{-5}+\Psi_0^{6,1}r^{-6}+O(r^{-7}),\label{chii0}\\
\psii_1&=&\Psi_1^{4,1}r^{-4}+O(r^{-5}), \label{chii1}\\
\psii_2&=&\Psi_2^{4,1}r^{-4}+O(r^{-5}), \label{chii2}\\
\psii_3&=&\Psi_3^{4,1}r^{-4}+O(r^{-5}), \label{chii3}\\
\psii_4&=&\Psi_4^{4,1}r^{-4}+O(r^{-5}). \label{chii4}
\end{eqnarray}
As discussed in the previous section, due to the presence of logarithmic terms in the connection one is required to construct the auxiliary field from a characteristic initial value problem. Initailly the coefficients in the expansion of the auxiliary field and those in the logarithmic terms of the original field will coincide. The higher terms in the expansions (order $1/r^7$ onwards) may differ at later times. However, the lower coefficients will always agree (including those tied to the Newman-Penrose constants!).  On the initial null hypersurface $\scrn_0$ one sets
\begin{equation}
\psii_0|_{\scrn_0}=\Psi_0^{5,1}|_{\scrn_0}+\Psi_0^{6,1}|_{\scrn_0}+\cdots.
\end{equation}
Note that this series expansion is bounded by that of $\Psi_0|_{\scrn0}$: $|\psii_0|_{\scrn_0}| \leq |\Psi_0|_{\scrn_0}|$. The remaining initial data is prescribed by setting $\psii_1|_{\scrz}=\ethbar\Psi_0^{4,0}$ and $\psii_2|_{\scrz}=\psii_3|_{\scrz}=\psii_4|_{\scrz}=0$ ($\scrz=\scrn_0\cap\scri^+$). The initial data on the initial null hypersurface $\scrn_0$ contains no logarithmic terms, however this will appear at later times due to the presence of $\ln r$ in the connection. It can be shown that these logarithmic terms due to the connection will appear at order $1/r^7$ from $\psii_0$ and $\psii_1$, and at order $1/r^6$ for the remaining components. Hence, they will not interfere with the evolution of the leading terms of the field. 

Direct calculations (similar to those in references \cite{javk98a} and \cite{javk99a}) show that the field given by the equations (\ref{chii0})-(\ref{chii4}) satisfies formally the spin-2 zero-rest-mass field equation $\nabla^{AA'}\psii_{ABCD}=0$, to any desired order. It has to be pointed out that the previous discussion is not a rigorous existence/uniqueness proof, but rather an argument in the formal expansions line of reference \cite{xiv}.

With regard to the conformal properties of the field $\psii_{ABCD}$, we will set 
\begin{equation}
\widehat{\psii}_{ABCD}=\Omega^{-1}\psii_{ABCD}.
\end{equation}
in order to keep the zero-rest-mass field equation conformally invariant. The leading components of the new auxiliary field will be of order $1/r^4$, hence from the Peeling theorem one can see that it will have no algebraically special part. The auxiliary field is not a radiation field.

On the same line of thought one can define another field that will contain all the non-logarithmic components of $\Psi_{ABCD}$:

\begin{eqnarray}
\psio_0&=&\Psi_0^{4,0}r^{-4}+\Psi_0^{5,0}r^{-5}+\Psi_0^{6,0}r^{-6}+O(r^{-7}),\\
\psio_1&=&\Psi_1^{4,0}r^{-4}+O(r^{-5}),\\
\psio_2&=&\Psi_2^{3,0}r^{-3}+O(r^{-4}),\\
\psio_3&=&\Psi_3^{2,0}r^{-2}+O(r^{-3}),\\
\psio_4&=&\Psi_4^{4,0}r^{-1}+O(r^{-2}).
\end{eqnarray}
If one demands that no logarithmic terms appear at the leading terms (e.g. at $1/r^4$ in $\psio_1$), then using the proposition 1, the field $\psio_{ABCD}$, will not satisfy the spin-2 zero rest-mass field  equations, but rather a non homogeneous version of it,

\begin{equation}
\nabla^{AA'}\psio_{ABCD}=f^{A'}_{BCD},
\end{equation}
the source terms coming from the terms with logarithms. 

If one wants to use the Kirchoff-d'Adhemar formula ``to evaluate the gravitational spin-2 field $\psi_{ABCD}$ at $i^+$'' (in the sense discussed in section 3) for a polyhomogeneous spacetime, one finds that the integrals diverge due to the non-regularity of $\psi_{ABCD}$ at null infinity. However, this `` evaluation at $i^+$'' is possible for the field $\psii_{ABCD}$ for it is regular at $\scri$.

As a result of the evaluation of $\psi_{ABCD}$ at $i^+$ one finds that the five conserved complex quantities

\begin{equation}
\oint_{\cals^2} \Psi_0^{6,1} (_2 \Ybar_{2,m}) \d \cals,
\end{equation}
are precisely the so-called \emph{logarithmic Newman-Penrose constants} of references \cite{xiv}, \cite{javk98a}, \cite{javk99a}. Using the arguments of Newman \& Penrose, it can be shown that these quantities are supertranslation invariant.

From this argument, it is clear why the quantities

\begin{equation}
\oint_{\cals^2} \Psi_0^{6,0} (_2 \Ybar_{2,m}) \d \cals,
\end{equation}
are not constants in the polyhomogeneous setting: the field associated with them ($\psio_{ABCD}$) is not a zero rest-mass field.

This discussion can be extended for general polyhomogeneous spacetimes, obtaining a hierarchy of fields $\stackrel{(N)}{\psi}_{ABCD}, \dots, \psio_{ABCD}$ in which only the field at the top of the hierarchy satisfies the spin-2 zero rest-mass field equation, while the other fields satisfy non-homogeneous versions, the source terms coming from the fields above in the hierarchical structure.

\section{Conclusions}
Some properties of spin-s massless fields have been studied. It has been shown how the notion of polyhomogeneity arises naturally when the fields into consideration do not peel. As discussed in the main text, an ``unhappy'' property of these non-peeling (polyhomogeneous) fields is that some of their tetrad components diverge at null infinity. To account for this problem, a hierarchy of auxiliary fields have been constructed. The fields in the hierarchy are esentially constructed out of the coefficients of the same degree in $\ln r$. One crucial property of the hierarchy of auxiliary fields is that the field at the top of it (the one built with the coefficients of the highest powers of $\ln r$) is also a zero-rest-mass field regular at $\scri$. This field is non radiative, and hence it contains information on the higher asymptotics of the field. In particular, one can construct the Newman-Penrose constants of the auxiliary field, that by construction are the same as the logarithmic NP constants of the original field. By means of this construction and the integral representation of the fields, one can interpret all the NP constants (logarithmic or not) as the value of a field at future infinity $i^+$. The remaining auxiliary fields of the hierarchy do not obey the zero-rest-mass equation, but a non-homogeneous version of it. This fact together with their divergence at $\scri$ precludes the use the Kirchoff-d'Adhemar for an evaluation at $i^+$. Hence they donot give rise to NP constants. Other possible uses of the present construction may include the discussion of the invariant transformations generated by the logarithmic constants (following the ideas of Glass \& Goldberg \cite{gg70}). This will be the subject of future research.

\section*{Acknowledgements}
I thank Prof. M.A.H. MacCallum for his friendly advice, suggestions and discussions. Thanks to Prof. D.C. Robinson for some ideas and for providing some nice references, and to Dr. R. Lazkoz and Dr. R. Vera for useful comments on the manuscript. I hold a scholarship (110441/110491) from the Consejo Nacional de Ciencia y Tecnolog\'{\i}a (CONACYT), Mexico.
\appendix

\section{Advanced and retarded linear fields}
Torrence \& Janis \cite{tj66} obtained set of advanced and retarded solutions for the spin-2 zero-rest-mass field equations in Minkowski spacetime. Their results can be easily extended to fields of arbitrary spin. The general results are listed here for reference.

Let,
\begin{equation}
K_p=\left( 2^p\frac{(l-p)!}{(l+p)!} \right)^{1/2}.
\end{equation}

\subsection{Advanced field for an arbitrary spin-$s$ zero rest-mass field.}

\begin{eqnarray}
\phi_0&=&K_{s}\sum_{l,m}(_sY_{l,m})r^{l-s}\partial_r^{l+s}\left(\frac{B^{lm}(v)}{r^{l-s+1}} \right), \\
& \vdots& \nonumber  \\
\phi_n&=&K_{s-n}\sum_{l,m}(_{s-n}Y_{l,m})r^{l-s}\partial_r^{l+s-n}\left(\frac{B^{lm}(v)}{r^{l-s+n+1}} \right),\\
 &\vdots& \nonumber \\
\phi_{2s}&=&K_{-s}\sum_{l,m}(_{-s}Y_{l,m})r^{l-s}\partial_r^{l-s}\left(\frac{B^{lm}(v)}{r^{l+s+1}} \right),
\end{eqnarray}
where $B^{lm}(v)$ is an arbitrary function of $v=u+2r$. These solutions are interpreted as advanced because they are constant on the $v=const.$ past null cones.

\subsection{Retarded field for an arbitrary spin-$s$ zero rest-mass field.}

\begin{eqnarray}
\phi_0&=&K_{s}\sum_{l,m}(_sY_{l,m})r^{l-s}(\partial_r-2\partial_u)^{l-s}\left(\frac{b^{lm}(u)}{r^{l+s+1}} \right), \\
& \vdots& \nonumber \\
\phi_n&=&K_{s-n}\sum_{l,m}(_{s-n}Y_{l,m})r^{l-s}(\partial_r-2\partial_u)^{l-s+n}\left(\frac{b^{lm}(u)}{r^{l+s-n+1}} \right),\\
 &\vdots& \nonumber \\
\phi_{2s}&=&K_{-s}\sum_{l,m}(_{-s}Y_{l,m})r^{l-s}(\partial_r-2\partial_u)^{l+s}\left(\frac{b^{lm}(u)}{r^{l-s+1}} \right),
\end{eqnarray}
where $b^{lm}(u)$ are arbitrary functions of the retarded time $u$. These functions are the time reversed counterpart of the advanced solutions listed above. Hence, they are interpreted as retarded radiation fields.

\section{The peeling spin-2 zero-rest-mass field in Minkowski spacetime.}

Setting $s=2$ in equations (\ref{NP1}) and (\ref{NP2}), one  obtains for \Minkowski \, the following set of equations:

\begin{eqnarray}
D\phi_n-\ethbar\phi_{n-1}&=&(2s-n+1)\rho\phi_n, \label{NPM1}\\
\Delta\phi_{n-1}-\eth\phi_n&=&-n\mu\phi_{n-1} \label{NPM2},
\end{eqnarray}
where $n=1,\ldots,4$. Assuming that the field peels, i.e.
\begin{equation}
\phi_n=\sum_{j=2s-n+1}\phi_n^jr^{-j},
\end{equation}
it is not difficult to solve the field equations (\ref{NPM1}) and (\ref{NPM2}) order by order. From the hypersurface equations (\ref{NPM1}) one gets:
\begin{eqnarray}
\phi_n^{6-n}&=&-\ethbar_1\phi_{n-1}^{6-n},  \\
\phi_n^{7-n}&=&-\frac{1}{2}\ethbar_1\phi_{n-1}^{7-n}, \nonumber \\
\phi_n^{8-n}&=&-\frac{1}{3}\ethbar_1\phi_{n-1}^{8-n}, \nonumber \\
 &\vdots& \nonumber
\end{eqnarray}
whereas from the evolution equations (\ref{NPM2}) one gets:
\begin{eqnarray}
\dot{\phi}_{n-1}^{6-n}&=&\eth_1\phi_n^{5-n}, \\
\dot{\phi}_{n-1}^{7-n}&=&\eth_1\phi_n^{6-n}-(3-n)\phi_{n-1}^{6-n}, \nonumber \\
\dot{\phi}_{n-1}^{8-n}&=&\eth_1\phi_n^{7-n}-(3-n+\frac{1}{2})\phi_{n-1}^{7-n}, \nonumber \\
 &\vdots& \nonumber
\end{eqnarray}

\section{Minimal no-peeling (polyhomogeneous) spin 2 zero-rest-mass fields in Minkowski spacetime.}
If the spin-2 zero-rest-mass field is minimally non-peeling, then
\begin{eqnarray}
\phi_0=\phi_0^4r^{-4}+\phi_0^5r^{-5}+\cdots, \\
\phi_n=\sum_{j=2s-n+1}\phi_n^jr^{-j},
\end{eqnarray}
where $n\geq2$, and the coefficients in the expansions are at most polynomials of degree 1 in $\ln r$. In particular one has:
\begin{equation}
\#\phi_0^4=\#\phi_2^3=\#\phi_3^2=\#\phi_3^3=\#\phi_4^1=\#\phi_4^2=\#\phi_4^3=0,
\end{equation}
i.e. the do not contain $\ln r$.
Solving the 4 hypersurfaces equations (\ref{NPM1}) one gets:
\begin{eqnarray}
\phi_1^{4\prime}&=&\ethbar_1\phi_0^4, \\
\phi_n^{6-n}&=&-\ethbar_1\phi_{n-1}^{6-n}+\phi_n^{6-n\prime}, \nonumber \\
\phi_n^{7-n}&=&-\frac{1}{2}\ethbar_1\phi_{n-1}^{7-n}+\phi_n^{7-n\prime}, \nonumber \\
\phi_n^{8-n}&=&-\frac{1}{3}\ethbar_1\phi_{n-1}^{8-n}+\phi_n^{8-n}\prime, \nonumber \\
 &\vdots& \nonumber
\end{eqnarray}
From the evolution equations (\ref{NPM2}):
\begin{eqnarray}
&&\dot{\phi}_0^4=0, \\
&&\dot{\phi}_0^5+\frac{3}{2}\phi_0^4-\eth_1\phi_1^4=0, \nonumber
\end{eqnarray}
and
\begin{eqnarray}
\dot{\phi}_{n-1}^{6-n}&=&\eth_1\phi_n^{5-n}, \\
\dot{\phi}_{n-1}^{7-n}&=&\eth_1\phi_n^{6-n}-(3-n)\phi_{n-1}^{6-n}+\frac{1}{2}\phi_{n-1}^{6-n\prime}, \nonumber \\
\dot{\phi}_{n-1}^{8-n}&=&\eth_1\phi_n^{7-n}-(3-n+\frac{1}{2})\phi_{n-1}^{7-n}+\frac{1}{2}\phi_{n-1}^{7-n\prime}, \nonumber \\
 &\vdots& \nonumber
\end{eqnarray}
for $n\geq 2$.

Note how the expressions one obtains for the non-peeling field are essentially the same ones for the peeling field, the only difference being the presence of terms of the form $\phi'$. These terms are all of them of degree zero in $ln r$, therefore the relations for the logarithmic coefficients are identical to those for the coefficients of the peeling field. This shows that the field $\phii_{ABCD}$ constructed out of the logarithmic coefficients in the way described in section 3, is a zero-rest-mass field. Generalizations to other classes of polyhomogeneous fields and different spins can be easily obtained.

\section{A minimal polyhomogeneous spin 2 zero-rest-mass field on a minimally polyhomogeneous background}

As a quick reference, the leading terms of the Weyl tensor, spin coefficients and tetrad functions for a  minimal polyhomogeneous spacetime are given:
\begin{eqnarray}
&&\Psi_0=\Psi_0^{4,0}r^{-4}+\left( \Psi_0^{5,1} \ln r +\Psi_0^{5,0} \right)r^{-5}+O(r^{-6}\ln r), \\
&&\Psi_1=\left( \Psi_1^{4,1}\ln r +\Psi_1^{4,0} \right)r^{-4} +O(r^{-5}\ln r), \\
&&\Psi_2=\Psi_2^{3,0}r^{-3}+O(r^{-4}\ln r), \\
&&\Psi_3=\Psi_3^{2,0}r^{-2}+\Psi_3^{3,0}r^{-3}+O(r^{-4}\ln r), \\
&&\Psi_4=\Psi_4^{1,0}r^{-1}+\Psi_4^{2,0}r^{-2}+\Psi_4^{3,0}r^{-3}+O(r^{-4}\ln r), 
\end{eqnarray}

\begin{eqnarray}
&&\rho=-r^{-1}-\sigma_{2,0}\sigmab_{2,0}r^{-3}+O(r^{-4} \ln r), \\
&&\sigma=\sigma_{2,0}r^{-2}+\sigma_{3,0}r^{-3}+O(r^{-4} \ln r), \\
&&\alpha=\alpha_{1,0}r^{-1}+\alpha_{2,0}r^{-2}+O(r^{-3} \ln r), \\
&&\beta=-\alphab_{1,0}r^{-1}+\beta_{2,0}r^{-2}+O(r^{-3} \ln r), \\
&&\tau=\tau_{2,0}r^{-2}+O(r^{-3}\ln r), \\
&&\pi=\pi_{2,0}r^{-2}+O(r^{-3}\ln r), \\
&&\gamma=\gamma_{2,0}r^{-2}+O(r^{-3}\ln r), \\
&&\mu= -1/2r^{-1}+\mu_{2,0}r^{-2}+O(r^{-3}\ln r), \\
&&\nu=\nu_{2,0}r^{-2}+O(r^{-3}\ln r),
\end{eqnarray}

\begin{eqnarray}
&&\xi^\alpha=\xi_0^\alpha r^{-1}-\sigma_{2,0}\xib_0^\alpha+O(r^{-3}\ln r), \\
&&C^\alpha=C_2^\alpha+O(r^{-3}\ln r), \\
&&Q=-1/2+Q_1r^{-1}+O(r^{-2}\ln r).
\end{eqnarray}

The hypersurface equations yield,

\begin{equation}
\phi_1^{4\prime}=\ethbar_1\phi_0^4 \left\{ 
 \begin{array}{l}
 \phi_1^{4,1}=\ethbar_1\phi_0^{4,0}
 \end{array}
\right.
\end{equation}

\begin{equation}
\phi_1^{5\prime}-\phi_1^5-\ethbar_1\phi_0^5-\ethbar_2\phi_0^4=\pi_2\phi_0^4 \left\{ 
 \begin{array}{l}
 -\phi_1^{5,1}-\ethbar_1\phi_0^{5,1}=0 \\
 \phi_1^{5,1}-\phi_1^{5,0}-\ethbar_1\phi_0^{5,0}-\ethbar_2\phi_0^{4,0}=\pi_{2,0}\phi_0^{4,0}
 \end{array}
\right.
\end{equation}

\begin{equation}
\phi_2^{4\prime}-\phi_2^4-\ethbar\phi_1^4=-\lambda_1\phi_0^4\left\{ 
 \begin{array}{l}
 -\phi_2^{4,1}-\ethbar\phi_1^{4,1}=-\lambda_{1,0}\phi_0^{4,0} \\
 \phi_2^{4,1}-\phi_2^{4,0}-\ethbar\phi_1^{4,0}=-\lambda_{1,0}\phi_0^{4,0}
 \end{array}
\right.
\end{equation}

\begin{equation}
\phi_3^{3\prime}-\phi_3^3=\ethbar_1\phi_2^3\left\{ 
 \begin{array}{l}
 -\phi_3^{3,1}=\ethbar_1\phi_2^{3,1} \\
 \phi_3^{3,1}-\phi_3^{3,0}=\ethbar_1\phi_2^{3,0}
 \end{array}
\right.
\end{equation}

\begin{equation}
\phi_4^{2\prime}-\phi_4^2=\ethbar_1\phi_3^2 \left\{ 
 \begin{array}{l}
 -\phi_4^{2,1}=\ethbar_1\phi_3^{2,1} \\
 \phi_4^{2,1}-\phi_4^{2,0}=\ethbar_1\phi_3^{2,0}
 \end{array}
\right.
\end{equation}

While the evolution equations give,

\begin{equation}
\dot{\phi}_0^4=0\left\{ 
 \begin{array}{l}
 \dot{\phi}_0^{4,0}=0
 \end{array}
\right.
\end{equation}

\begin{equation}
\dot{\phi}_0^5+\frac{3}{2}\phi_0^4-\eth_1\phi_1^4=3\sigma_2\phi_2^3 \left\{ 
 \begin{array}{l}
 \dot{\phi}_0^{5,1}-\eth_1\phi_1^{4,1}=0 \\
 \dot{\phi}_0^{5,0}+\frac{3}{2}\phi_0^{4,0}-\eth_1\phi_1^{4,0}=3\sigma_{2,0}\phi_2^{3,0}
 \end{array}
\right.
\end{equation}

\begin{equation}
\dot{\phi}_1^4-\eth_1\phi_2^3=2\sigma_2\phi_3^2 \left\{ 
 \begin{array}{l}
 \dot{\phi}_1^{4,1}=0 \\
 \dot{\phi}_1^{4,0}-\eth_1\phi_2^{3,0}=2\sigma_{2,0}\phi_3^{2,0}
 \end{array}
\right.
\end{equation}

\begin{equation}
\dot{\phi}_2^3-\eth_1\phi_3^2=\sigma_2\phi_4^1 \left\{ 
 \begin{array}{l}
 \dot{\phi}_2^{3,0}-\eth_1\phi_3^{2,0}=\sigma_{2,0}\phi_4^{1,0}
 \end{array}
\right.
\end{equation}

\begin{equation}
\dot{\phi}_3^2-\eth_1\phi_4^1=0 \left\{ 
 \begin{array}{l}
 \dot{\phi}_3^{2,0}-\eth_1\phi_4^{1,0}=0
 \end{array}
\right.
\end{equation}

The hypersurface and evolution equations for the leading terms of the auxiliary field are the same ones of the coefficients of the form $\phi_n^{x,1}$.

\end{document}